\newif\ifanonymized
\anonymizedfalse    

\newif\ifsubmit
\submittrue	

\documentclass[acmsmall]{acmart}
\usepackage{longtable}

\usepackage{fancyhdr}

\usepackage{url}

\usepackage{color}

\usepackage[mmddyyyy,hhmmss]{datetime}

\usepackage{xspace}

\usepackage{graphicx}
\DeclareGraphicsExtensions{.pdf,.png,.eps,.ps,.jpg}

\ifsubmit
  \newcommand{\hey}[1]{\relax}
  \newcommand{\heyvarun}[1]{\relax}
  \newcommand{\bibtex}[1]{\relax}
  \newcommand{\note}[1]{\relax}
\else
  \newcommand{\hey}[1]{\textcolor{magenta}{[{#1}]}}
  \newcommand{\heyvarun}[1]{\textcolor{blue}{[Varun: {#1}]}}
  \newcommand{\note}[1]{\par\textcolor{magenta}{Note: {#1}}\par}
  \newcommand{\bibtex}[1]{\textcolor{red}{@bibtex}\{#1\}} 
\fi

\newcommand{\hide}[1]{\relax}

\newcommand{\seclabel}[1]{\label{sec:#1}}

\begin{document}

\title{Navigating the Paradox: Challenges and Strategies of University Students Managing Mental Health Medication in Real-World Practices}
\date{}


\ifanonymized
   \author{Anonymized for blind submission}
\else
   \author{Jiachen Li}
   \affiliation{
    \institution{Northeastern University}
    \city{Boston}
    \state{MA}
    \country{USA}}
    \email{li.jiachen4@northeastern.edu}
    \author{Justin Steinberg}
   \affiliation{
    \institution{Northeastern University}
    \city{Boston}
    \state{MA}
    \country{USA}}
    \email{steinberg.ju@northeastern.edu}
     \author{Elizabeth Mynatt}
   \affiliation{
    \institution{Northeastern University}
    \city{Boston}
    \state{MA}
    \country{USA}}
    \email{e.mynatt@northeastern.edu}
    \author{Varun Mishra}
   \affiliation{
    \institution{Northeastern University}
    \city{Boston}
    \state{MA}
    \country{USA}}
    \email{v.mishra@northeastern.edu}
    
\fi



\begin{abstract}
Mental health has become a growing concern among university students. While medication is a common treatment, understanding how university students manage their medication for mental health symptoms in real-world practice has not been fully explored. In this study, we conducted semi-structured interviews with university students to understand the unique challenges in the mental health medication management process and their coping strategies, particularly examining the role of various technologies in this process. We discovered that due to struggles with self-acceptance and the interdependent relationship between medication, symptoms, schedules, and life changes, the medication management process for students was a highly dynamic journey involving frequent dosage changes. Thus, students adopted flexible strategies of using minimal technology to manage their medication in different situations while maintaining a high degree of autonomy. Based on our findings, we propose design implications for future technologies to seamlessly integrate into their daily lives and assist students in managing their mental health medications.
\end{abstract}
\maketitle  




\ifsubmit
    \relax
\else
    \par\noindent \textcolor{red}{\textbf{DRAFT}: \today\ -- \currenttime}
    \pagestyle{fancy}
    \lhead{DRAFT in preparation}
    \rhead{version: \today\ -- \currenttime}
    \chead{}
    \lfoot{}
    \rfoot{}
\fi
\definecolor{DarkGreen}{HTML}{000000}
\newcommand\review[1]{\textcolor{DarkGreen}{#1}}
\section{Introduction} 
\seclabel{Introduction}

Researchers have recently observed that mental health challenges pose a significant concern, particularly among university students. According to the Healthy Minds Study (HMS) National Report 2022-2023, 46\% of students have been diagnosed with a mental disorder by a health professional, encompassing depression (30\%), anxiety (36\%), and other symptoms~\cite{network2023healthy}. The reported data represents the highest rate in the 15-year history of the survey, reflecting a nearly 50\% increase from 2013 to 2021~\cite{lipson2022trends}. Recognizing the severity of these issues, many researchers have developed various technologies to assist, including chatbots~\cite{de2020investigating,fitzpatrick2017delivering,info:doi/10.2196/10148}, digital therapeutic apps~\cite{richards2015beyond,bauer2018acceptability}, and wearables for passive monitoring~\cite{10.1145/3351274,10.1145/3341162.3349304,mishra:stress-reproducibility}. Although these technologies have demonstrated positive effects, one significant aspect of the mental health management journey is often understudied -- the role of mental health medication. According to the report, 40\% of students with positive depression or anxiety screens had taken at least one type of medication in the preceding 12 months. Since these medications often have a significant effect in mitigating and managing mental health symptoms, it is crucial to examine people's experiences in taking and managing these medications.

When contemplating medication management strategies for mental health issues, we recognize that it is more intricate compared to some other diseases, primarily due to the complex relationship between mental health symptoms and medication effects. While medication issuance and dosage adjustments for many diseases are typically grounded in test results, diagnosing a mental health condition necessitates profound introspection into internal states, such as mood, which can be more `subjective.' Consequently, addressing medication for mental health involves a greater reliance on individuals' self-reported behaviors and experiences than objective test results, implying that the issuance and adjustment of medication could be dynamic and unstable. Furthermore, managing medication itself might influence or be influenced by the person's mental status, which cannot be merely regarded as an external factor for medication non-adherence but also as symptoms the medications seek to treat. This intricate connection between symptoms and treatment introduces unique complexity in managing medications for mental health concerns. In this paper, we seek to gain insights into students' real-world experiences and challenges in managing their mental health medications.

Understanding mental health medication management strategies for college students is relatively understudied. Most prior works on medication management strategies focused on older adults~\cite{conn2009interventions,yap2016systematic} and individuals with disabilities~\cite{schwartz2021inclusion}, often categorizing younger populations as `healthier.' With the increase in mental health diagnoses, however, more college students are regularly taking medications to manage their mental health. Further, given their busy schedules and dynamic routines, ensuring adherence among this group is also difficult. For instance, Gray et al. discovered that students adhered to only 53\% of prescribed ADHD medication doses, a rate associated with the varying times throughout the semester~\cite{benson2015misuse}. For researchers aiming to develop technologies to address this issue, it is important to first understand the real-world practices of individuals in adopting various technologies for mental health medication management, the challenges they encounter during the process, and the hesitations with being more consistent with the medications. Recognizing the significance of this issue, our research delved into the challenges university students face in managing their mental health medication and the coping strategies they employ to tackle these issues.

\medskip
\noindent
In this paper, we make three important \textbf{contributions}: 
\begin{itemize}
    \item We pinpoint the unique challenges in the medication management process for university students with mental health conditions.
    \item We delve into the coping strategies employed by university students in addressing these challenges, with a particular focus on examining the role of various technologies in the process.
    \item Based on our findings, we outline and discuss design implications for future systems aimed at assisting university students in managing their mental health medications.
\end{itemize}

\section{Related Work} 
\seclabel{related}

\subsection{Mental health issues among university students}
The prevalence of mental health issues among university students has been escalating, posing a growing concern. According to a recent study, nearly half of all lifetime mental disorders start by the mid-teens and three-quarters by the mid-20s ~\cite{kessler2007age}. The discussion of mental health within college communities underscores the challenges that many students face during their collegiate journeys. The combination of heavy academic workloads, social pressures, and the inherent uncertainties associated with this exploratory phase of life contributes to an environment where mental health issues can manifest ~\cite{pedrelli2015college,kitzrow2003mental,zivin2009persistence}.

In recent years, there has been a discernible uptick in mental health concerns among students, with a notable rise in conditions such as anxiety, depression, eating disorders, and suicidal ideation ~\cite{pedrelli2015college,network2023healthy}. These issues not only cast shadows over the overall college experience but also impede students' active participation in academic and extracurricular activities, leading to a considerable decline in their overall quality of life ~\cite{doran2021prescribing, lipson2022trends}. Due to the gravity of the situation, there has been a stronger push to spread mental health awareness. Between 2013 and 2023, there has been a substantial 7.8\% reduction in the perceived stigma surrounding mental health issues, fostering an environment where individuals are more inclined to seek assistance ~\cite{lipson2022trends, network2023healthy}. This reduction in stigma has correlated with a 12.8\% increase in the utilization of psychotropic medications among college students during the same period ~\cite{lipson2022trends}.

While the surge in medication use reflects a growing acknowledgment of mental health challenges, the issue of mental medication adherence among university students remains under-explored. The efficacy of mental health therapy and psychiatric medication is contingent on consistent adherence to prescribed regimens. Factors such as academic pressures, social dynamics, and lifestyle transitions unique to the college experience may significantly influence medication adherence\review{~\cite{ashouri2015review,wagoner2020barriers,shahin2021association,wagoner2020barriers}}, thereby necessitating a nuanced understanding of this critical aspect of mental health treatment.
\subsection{Mental health treatment: the role of medications}

The treatment of mental illness frequently involves a combination of medication and mental therapy. Medications are typically prescribed when psychotherapies on their own are ineffective ~\cite{doran2021prescribing, CHOU201669}. Medications play a crucial role in alleviating symptoms by influencing brain chemicals affecting mood and behavior ~\cite{bushnell2018treating}. Commonly prescribed medications include antidepressants, anti-psychotics, and mood stabilizers ~\cite{merikangas2013medication}. Antidepressants are used to treat a wide variety of illnesses, such as major depressive disorder (MDD), general anxiety disorder (GAD), and obsessive-compulsive disorder (OCD) ~\cite{LEPPIEN202013}. When taking antidepressants, the dosage of the medication will help determine the effectiveness of the treatment. However, the higher the dose, the more likely a subject will experience side effects ~\cite{doi:10.1056/NEJMoa0804633}. 

While psychiatric medications are integral components of mental health treatment, their effectiveness is also contingent on factors such as accurate diagnosis, tailored prescription regimens, and, notably, adherence to prescribed medications~\cite{mahmud2023effectiveness, babel2021artificial, MARRERO20202116}. Adherence is the execution of a prescribed treatment plan with little to no error. This includes taking the correct medication at the appropriate dose, at the proper time, and with a proper frequency ~\cite{jimmy2011patient}. Adherence to psychiatric medication regimens is a critical determinant of treatment success, influencing the ability to manage symptoms and enhance overall functioning ~\cite{MU2023103766, ROOTESMURDY2018613}. There are many factors that can influence a subject's capacity to adhere to medication, such as age, educational level, attitudes and beliefs towards medication, \review{routines and schedule, stigma, dosage changes,} and medication side-effects \review{~\cite{MARRERO20202116,wagoner2020barriers,papoutsaki2021understanding, caldeira2022state}}. We contribute to the existing literature by understanding how these factors affect adherence within the college student demographic, specifically for mental health medications.

\subsection{Technology for medication management}

In order to combat the issue of medication non-adherence, many studies have tested different types of adherence-related technologies. There is no universally recognized standard in medication adherence technology, making it difficult for patients to identify the most effective solution for their needs. These technologies encompass a wide range of tools such as smart pill bottles, sensors, self-reporting technology, mobile applications, and chatbots \review{~\cite{stawarz2014don, doi:10.1177/0193945914524492, PAL2021183, 10.1093/jamiaopen/ooab021}}. Smart pill bottles, for instance, can connect to the internet and record the time when a patient opens the bottle, providing a timeline of adherence ~\cite{RICE2021117612}. For mobile phones, there are many different types of modalities that can be utilized, such as phone calls, short messaging service (SMS) texting, online surveys, as well as audio and video content \review{~\cite{mason2022technologies, doi:10.1177/0193945914524492, APPIAH2021100005,stawarz2014don}}. However, depending on the utilized feature(s), there can be challenges with user engagement, battery life, storage capacity, and ease of use ~\cite{mason2022technologies}. 

Notably, the combination of mobile interventions with smart pillboxes has emerged as an effective method in supporting antidepressant treatment ~\cite{doi:10.1177/2055207616663069}. Mobile applications, in this context, serve not only to monitor adherence but also to provide medication education, track side effects, and symptoms, and act as a communication channel with healthcare providers. This serves to address multiple pain points that are related to adherence, such as a lack of knowledge, forgetfulness, and low motivation ~\cite{doi:10.1177/2055207616663069}. This multi-layered approach showcases the potential of technology to address medication adherence challenges while offering additional benefits in terms of education, monitoring, and communication within the healthcare ecosystem. 

The exploration of using mobile adherence solutions also extends to Artificial Intelligence (AI). AI smartphone applications have also been investigated for their effectiveness in delivering reminders and encouragement for medication adherence for adults with Non-communicable diseases (NCD) ~\cite{babel2021artificial}. Conversational agents, also known as chatbots, are able to simulate human interaction in text, audio, or visual formats ~\cite{info:doi/10.2196/40789}. Conversational AI systems have been shown to help older adults remember to take their medications by tracking their daily activities and symptoms ~\cite{babel2021artificial}. AI systems are able to collect and interpret data from a wide variety of different data structures, such as from electronic health records, clinical notes, and real-time physiological data in conjunction with sensors \cite{info:doi/10.2196/40789}. Furthermore, because of an AI system's ability to detect meaningful patterns in data, they also have the ability to provide personalized responses to better support user needs as well as foster patient/provider relationships ~\cite{babel2021artificial, info:doi/10.2196/40789}. In addition, the integration of these new technology-based approaches comes with several advantages, including easy implementation, cost-effectiveness, and extensive outreach ~\cite{ROOTESMURDY2018613, mahmud2023effectiveness}. Our research focuses on discerning effective versus ineffective technologies for medication adherence among college students and understanding how technology-specific factors influence adoption.   
\section{Method} 
\seclabel{method}
\subsection{Procedure}
\review{Recruitment materials were distributed to participants within various university networks. Participants were required to provide a '.edu' email to verify their student status. After expressing initial interest, w}e emailed the consent form to the participants, outlining the study's purpose and content to prepare them for potentially sensitive interview topics before the interview. If the participants wanted to proceed, we scheduled the virtual semi-structured interviews \review{through Zoom}, which typically lasted an hour. Two members of the research team were present for all interviews; one researcher led the interview while the other documented the discussion. We obtained audio recordings and transcripts only after the participants' initial consent. \review{Each participant received a \$20 Amazon gift card via email after the interview.}

The interviews primarily revolved around each participant's unique journey managing their mental health medication. We delved into their medication schedules, ranging from basic details like frequency to deeper explorations of personal challenges. Our focus also extended to the technologies used during this process, exploring both successful and challenging experiences. Acknowledging the individuality of each journey, we let the participant drive the interview and describe their experience in the later part of the interview\review{, with some guiding prompts like, \emph{"Describe your most recent experience taking medication, such as today or yesterday."}} Researchers dynamically adjusted the interview questions based on each participant's unique challenges, aiming to capture a comprehensive understanding of their personal experiences in managing medications within real-world scenarios.

In addition to the interviews, we requested the participants to complete an online survey. This survey collected demographic information, basic medication information and three scales measuring adherence and self-efficacy in medication use: the Self-Efficacy for Appropriate Medication Use Scale (SEAMS)~\cite{risser2007development}, Morisky Medication Adherence Scale (MMAS-8)~\cite{morisky2008predictive} \review{that have been validated and used in multiple studies~\cite{lam2015medication}}, and a self-designed survey \review{[Table. 3]}.

\subsection{Participants}
In this study, we conducted ten semi-structured interviews with participants who are self-identified as currently taking medication for mental health diagnosis and are students at a university. We used the participants' university email addresses to verify their eligibility for the study. 

Two of our participants identified as male, while eight identified as female. In terms of racial identification, five participants identified as White, one as Hispanic, one as Black or African American, one as Asian, and two as Other. Among the participants, three were enrolled in undergraduate school, and the remaining seven were attending graduate school. On average, participants reported taking 1.6 medications (SD=0.70). All participants took medications at least once a day, with one individual taking additional medications weekly and another multiple times a day. Regarding medication storage, eight out of ten participants only used the original pill bottle or blister; one used a 7-day drug store-style pillbox, and another used both methods. In terms of technology utilized in the medication management process, four relied solely on memory, four used general digital calendar/reminder apps, and two utilized other technologies such as third-party medication and self-report apps.

The average scores for medication self-efficacy and adherence were 25.3 (scale value ranging from 13 as non-efficacious to 39 as efficacious) and 4.2 (scale value ranging from 0 as adherent to 7 as non-adherent), respectively. These scores suggest that there is no obvious ceiling effect in our sample, indicating that participants do face some challenges in their medication management. We present detailed demographic information for the participants in the appendix.


\subsection{Data Analysis}
We carefully reviewed the interview transcripts and notes, adopting an open coding approach~\cite{corbin1990grounded}. The data were coded independently by two researchers, with each focusing on specific segments. Subsequently, both researchers exchanged their individual codebooks, engaging in discussions to harmonize and merge the codes \review{by consensus}. Following this collaborative process, we conducted a thematic analysis, adhering to the principles of Grounded Theory~\cite{braun2006using, glaser1968discovery}. \review{We reached data saturation with 10 participants and stopped further recruitment~\cite{Guest2006HowMI,Ando2014AchievingSI}.} This analysis resulted in the identification of three primary themes we detail in the findings section. 

\subsection{Ethical consideration}
The study obtained IRB approval from the research institution. Given the sensitive nature of the data collected, personal identification was immediately removed post-interview. Participants were given the autonomy to self-report their diagnosis voluntarily and retained the right to withhold this information if they chose. \review{Realizing the interview topic's sensitivity and the potential risk of disclosing private information, we ensured that recruitment of new participants ceased as soon as we reached data saturation~\cite{Guest2006HowMI,Ando2014AchievingSI}. } 
\section{Findings} 
\seclabel{findings}
In this section, we describe our key findings from the thematic analysis.
\subsection{A dynamic voyage: Managing mental health medication is never a static process}

A standout takeaway from all the interviews was the dynamic nature of managing mental health medication. In this section, we aim to illustrate the causes of this dynamic nature and its effects on medication management strategies.

\subsubsection{Struggling with Self-Acceptance: Expectations of Reducing or Ceasing Medication}
We noticed significant concern and hesitation regarding taking medication for mental health concerns, setting apart the expectations for mental health medication from those for treating other chronic diseases. 9 out of 10 participants did not commence medication immediately after receiving a diagnosis. For instance, P5 had been diagnosed with a mental health condition for 16 years but only began taking medication approximately 10 years ago. This delay in medication initiation underscores that medication isn't typically the first treatment choice for mental health challenges. 
\begin{quote}
    \textit{"I just wasn't seeking medical help like medical or medicine intervention for a while." (P4)}
\end{quote}
Starting medication was often considered \textit{`a tough decision'} by our participants, generating feelings of anxiety. This anxiety stemmed from various reasons. Initially, many associated taking medication with \textit{`having serious issues'} and struggled to reconcile this notion within themselves. 
\begin{quote}
    \textit{"For example, once you start going on medication, it's like, you have problems. It's more like admitting to yourself that you need help." (P2)}
\end{quote}
Moreover, uncertainty about the effects of medication caused anxiety. P2 described embarking on taking medications as \textit{`taking a risk'} since it's not clear why some people have different symptoms even on the same medications.

Concerns also extended to potential long-term effects, particularly as mental health medication \textit{`has some influence on our brain'} (P8). Another intriguing factor was the involvement of parents in this decision-making process, especially considering the changing relationships between university students and their parents during this pivotal time. Participants mentioned instances where their parents decided if they could take medication during their teenage years, or they felt the need for parental confirmation before starting medication in college while gradually assuming greater autonomy later on. We delve deeper into this transitional period in the latter part of this paper.
 
Due to concerns about taking mental health medication, although many patients ended up using it for a long time, they did not expect nor desire a long-term reliance on medications. P4 and many others expressed that they didn't anticipate medication being a lifelong necessity; instead, they hoped to gradually reduce and eventually stop taking it. P8 expressed strong feelings on this:
\begin{quote}
    \textit{"I want to decrease overall medication usage. And I really hope that one day I will not rely on the medication." (P8)}
\end{quote}
 This unique desire leads to more instability in managing mental health medication, unlike managing many chronic diseases where patients are comfortable maintaining a stable dose for extended periods~\cite{coleman2012dosing}. 
 Our participants consistently reevaluate their symptoms, actively seeking opportunities to adjust the dosage. This frequent reassessment significantly destabilizes the management of medication.

\subsubsection{Dosage changes and medication adjustments are common}

In addition to the participant's willingness to reduce dosage, several other factors in real life may contribute to changes in dosage or even the type of medication. A common cause for adjusting dosage is the fluctuation in symptom severity. This occurrence is surprisingly frequent among our participants, given that mental health symptoms can be significantly and rapidly influenced by life changes, a common experience for university students. P10 increased dosage after entering college to manage the heightened stress of schoolwork, while P8 found a better balance and relied less on medication upon studying abroad for grad school. 

Insufficient effectiveness and significant side effects are substantial factors leading participants to alter their dosage or medication. We observed a noteworthy number of participants expressing the challenge of finding the right medication that works for them, particularly at the onset of their medication journey. P4 described this process in detail:
\begin{quote}
    \textit{"You feel elaborate when you have to figure out what works for you. If you visit with a professional like a psychiatrist, and they prescribe something to the best of their knowledge. You have to stay on it for like 2 months to see the full effects, which means if it's something [that] doesn't work for you, you have to commit to it for 2 months, and then go back to the drawing board. Sometimes you have to wean off of it. It's like, hmm, start over." (P4)}
\end{quote}
 P1 also described people underestimating the risks when starting a mental health medication:
 \begin{quote}
    \textit{"I know people expect sometimes it has an effect right away. I think it's not commonly known that within the first couple of weeks of taking medication are actually, if you're at risk for suicide, those are particularly volatile weeks for you in starting a medication, because it's a frequent occurrence that starting a medication can actually give people a sort of boost of motivation, almost that they need to follow through with any sort of suicidal ideation or self-harm." (P1)}
\end{quote}
This revealed a gap between theory and real-world experiences of taking medication, where there's more back and forth, which would complicate medication management strategies.
To summarize, the previous descriptions explain why there are frequent changes in medication for university students with mental health issues, emphasizing the importance of recognizing and addressing these issues.

\subsubsection{Medication interlaces with mental health symptoms}

Aside from treating symptoms, medication for mental health establishes a complex and intimate relationship with mental health status. While many of our participants recognized the importance of taking medication on time, they also found this additional task to be stressful, which in turn deteriorated the symptoms. 

\begin{quote}
    \textit{"I also constantly feel the pressure of stress which reminds me to take the medication that has been prescribed for that stress." (P1)}
\end{quote}
 P4 also found this process distressing \textit{"because you're altering your mental state in an attempt to fix your mental state."} Mental health symptoms also impact participants' capacity to take medication that alleviates those symptoms. For instance, P5 mentioned that during particularly depressive episodes, getting out of bed and accessing medication stored in the kitchen became challenging. As a result, the distinctive connection between mental health symptoms and managing medication designed to address those symptoms further complicates the process.

\subsubsection{Transition period for university students}
Besides the instability caused by medication and mental symptoms, the unique phase of being university students introduced further fluctuations in the medication management process. Many participants indicated significant changes in their medication routines upon entering school. This involved practical challenges such as difference in accessing medical services due to relocation and changes in insurance (typically transitioning from family to individual coverage). Moreover, this life stage shift represented a change in responsibility, prompting university students to manage their medication independently. As P1 noted:
\begin{quote}
    \textit{"My mom would remind me a lot when I was at home. Almost daily or just even ask me if I had taken it, so I definitely lack that (at school)." (P1)}
\end{quote} 

Participants also expressed that the decision to take medication was no longer solely their parents' responsibility, granting them more autonomy in decision-making. Predictably, this life transition resulted in unstable medication management strategies as university students adapted to a new lifestyle and worked on establishing new habits.

In this section, we discovered that the process of taking mental health medication was marked by instability and continual fluctuations, influenced by various factors related to the self, medication, and external events. This inherent instability prompted university students to develop flexible strategies for managing their medication, a topic we will delve into in detail in the next section.

\subsection{Strategies for managing mental health medication among university students: Flexible and Creative Ways of Using Simple Technology}

This section reflects the diversity of needs and corresponding methods college students use to better incorporate medication management into their dynamic lifestyles. Before exploring different strategies, it's crucial to delve into the motivation and consequences behind university students taking medication. This context is essential for understanding why they opt for specific strategies.
\subsubsection{Motivation and consequences: Symptoms}
A large reason why our participants have developed or adopted strategies in the first place is due to the consequence of side effects or decline in mental health and/or performance that comes with missing dosages. Many of the consequences are experienced firsthand while the motivation of others is driven by worry or anxiety. According to P7, 
\begin{quote} \textit{“If not taking my meds means that I feel horrible, or I have a depressive episode, or can’t get the work done, or I don’t show up in the way I want to enough times, I am motivated to take my meds on time. If I can’t feel the difference, I’ll spontaneously stop taking them, because I don’t think they’re making a difference... I think it’s rarely me being motivated to take my meds. It’s me being motivated to not feel like s**t.”}
\end{quote}
Further, due to the nature of many mental health medications, missing doses do not have an immediate effect on the user, which can cause the user to become more relaxed with their routine. Nevertheless, all of our participants recognized the importance of \textit{"being more consistent"} and generally experienced heightened stress when missing a dose for an extended period, which could lead to a severe outcome. P2 once took their own initiative to quit medication after forgetting to take medications a few days, and \textit{“was just kinda like, f**k it, I'm fine.”}. However, this did not yield positive results, and P2 had to resume their medication after a while when they felt the symptoms returning. This nicely characterizes why our participants find it so important to use different tools and strategies to aid them in their medication adherence.

\subsubsection{Fixed reminders are common but not enough}

Many participants developed strategies involving alarms and reminders as parts of their daily routines. Six out of 10 participants used the pre-installed iPhone clock application due to its easy access, and only a minority (2) sought out specialized apps such as Apple Health. Despite the use of various applications, all reminders were set at fixed times, typically aligned with their regular medication schedule. However, participants expressed that this approach is often insufficient for various reasons.

Firstly, our participants sometimes don't immediately take their medication when the alarm goes off. For example,
\begin{quote}
    \textit{"If I'm just chilling at that point I'll just take my pill. If I am wrapping something up, like I'm playing a video game or watching a show and I have 5 more minutes left, I just gotta finish this thing. (P2)"}
\end{quote}
Depending on their current activities, some individuals may choose to snooze the alarm if they are in the middle of a task. Moreover, if there is a longer delay before taking the medication, participants dynamically adjust the reminder time. For example, if P2 is not at home at 9 pm (the usual time of the recurring alarm), they turn off the 9 pm alarm and manually set a new alarm for 11:30 pm to be reminded later. They then repeat the snooze method until they have taken their medication. Some participants even employ a backup reminder strategy at the beginning, setting two recurring alarms at different times in case they miss the first one.

Secondly, their medication routines might be \textbf{closely tied to other schedules that do not adhere to fixed times}. P5's perspective highlights the significance of a system that acknowledges the variability in waking times of college students and emphasizes the need for reminders to adapt accordingly.
\begin{quote}
    \textit{"I wake up at not exactly the same time every day, and my medication routine is tied to my morning routine rather than a specific time. So setting an alarm that I need to take it at 9 am or 10 am every day isn't as useful for me(P5)"}
\end{quote}
In addition to the sleep schedule, P3 also mentioned that their medications \textit{"don't do very well with skipping breakfast."}. This implies that having a meal is another crucial activity that participants need to consider when managing their medication.

The earlier discussions on schedules illustrate how routine activities related to medications occur at flexible times. Moreover, the situation could worsen if our participants have an \textbf{irregular schedule} on some days that disrupts their normal plan. For instance,
\begin{quote}
    \textit{"If I wake up sort of later in the day, and there's a lot that I need to do, I'm more likely to skip that normal morning routine that I stick with, which tends to just sort of push my medication to the back of my mind." (P1)}
\end{quote}
Academic tasks such as meetings are sometimes events that often get in the way. In those cases, taking medication won't be prioritized, and \textit{"sometimes I just end up forgetting because I don't have the allowance or the reminders anymore"} (P3). In more complicated situations where P5's medications were linked with meals, they missed a dose because the 7am meeting was earlier than their usual wakeup time, disabling them to take breakfast which resulted in forgetting medications.

Due to the limitations of fixed alarms, some participants did not consider reminders and alarms to be the prioritized strategy, but rather viewed them as a double assurance. For P9, they had a daily reminder but still usually took medications before the alarm.

To summarize, this perspective calls for a shift from a rigid, time-based reminder to a more adaptive and context-aware system. Participants expressed desire for technology that understands and integrates into their established daily routines. This ensures that medication reminders are not only timely, but align with their natural routines.

\subsubsection{Physical cues: A simple but effective method}
While acknowledging the limitations of digital reminders, several participants in our study expressed a preference for using physical cues to manage their medication. For instance, P7 highlighted their reliance on a physical pill organizer right next to their bed. Similarly, P10 strategically places their medications on their desk near their computer and backpacks. By keeping the pillbox visible and within their line of sight, they create visual cues that serve as reminders for medication intake. The preference for physical instruments is rooted in a sense of reliability and simplicity. This insight sheds light on the significance of environmental cues and the impact of visibility on adherence behavior. 

\subsubsection{Gaps between reminder and actual intake: The paradox of logging}
While reminders are helpful in medication management, there's still a gap between setting up the reminders and actually taking the medications on time. P3 shared a sentiment about how easy it is to become distracted and cast doubt upon oneself regarding whether they have actually taken their medication. When the alarm goes off, they have to walk over to the place where they keep their medication. If they are thinking about something else at that specific moment, they can forget whether they have taken it. P2 also shared a strategic of turning the phone alarm off after but not before taking medications to avoid uncertainty caused by distractions.
In such cases, proper tracking and logs for medication become crucial. While a minority of our participants were logging their medication intakes, the majority of them (9 out of 10) were not doing so, even if they recognized the possibility and importance. This did not exactly align with our expectations of university students, who are generally more tech-savvy. When discussing the requirements for a future system to assist with medication management, the most commonly mentioned keywords were minimal and simple. They don't want to \textit{"add an additional task"} to their established routines, with self-reporting typically being considered one of those tasks. P4 complained:
\begin{quote}
    \textit{"I'm like, not great at logging symptoms like, not great, basically just any longitudinal logging, like, I just don't love doing it. "}
\end{quote}
The need for logging was not just for medication but also for symptoms related to medication, such as moods and side effects. 
Similar to Section 4.1.3, participants express a desire to monitor their mood and medication dosage to deduce a potential relationship, given the intricate connections between symptoms and medications. However, these efforts often face challenges, primarily due to the complexity and additional effort required for ongoing self-reporting. Here, we observed a paradox between the necessity of having this information and the reluctance to put in extra effort to acquire it.

\subsubsection{Preserving autonomy in medication management: Managed by symptoms}
In addition to finding ways to ensure they take medication on time, we also realized that our participants might occasionally decide to skip medication for various reasons. Among those reasons, an interesting factor was mental health symptoms. Because the primary motivation for taking medication is to treat the symptoms, some participants also \textbf{manage medication based on symptoms}. P4 mentioned they never forgot to take medication for more than 3 days because, in that case, they would experience symptoms that reminded them to take their medications.
\begin{quote}
    \textit{"Is there a reason why today's harder than other days? Is there a reason why I'm doing more work to take care of myself mentally than other days? And then usually it'll be like, did I take my meds? No, I didn't, oh, shoot!"}
\end{quote}

In some cases, our participants would even take medication immediately once they felt the symptoms, usually for short-term effect meds. For instance, one of P5's medications has a short-term effect. They expressed themselves not having a good routine and take medication immediately if they felt the symptoms, but would skip the next dose if it was too close.

All of these reflect participants' urgent needs to preserve autonomy in managing medication. That's part of the reason why, even if they acknowledge the limitations of oversimplified technology such as default alarms, they still chose to use it and adapt it in their own ways so that they can have full control over it. 

\subsection{Overcoming obstacles in seeking external support}
Besides coping with personal concerns, motivation, and the inconvenience of managing medication independently, our participants encountered numerous obstacles while seeking external support from various stakeholders. One of the most common barriers involved the difficulty in \textbf{obtaining medication}. In our study, all mental health medications were prescribed by doctors, necessitating patients to find a doctor, obtain a diagnosis, and then acquire a prescription. Many participants found this process overwhelming, primarily due to the multiple rounds of interactions with different professionals, particularly as most medications required issuance by specialists, such as psychiatrists. \textit{"It just ended up being a little bit too many phone calls and a little bit too much like I didn't have enough understanding on how,"} expressed one participant, who eventually gave up seeking medication due to the complex procedures on the first try. P4 highlighted the frustration of starting the process repeatedly because of frequent relocations for school and changes in general practitioners. Additionally, P3 and P8, who were international students, mentioned the extended difficulty in transferring prescriptions due to the US medical system's inability to validate previous diagnoses and prescriptions. Despite having local student insurance, P8 continued to consult their doctors remotely in their home country and shipped the medication overseas. Some participants initially sought help from school medical centers but often faced long wait times. Another challenge surfaced with obtaining stimulant medication for ADHD, which faced shortages and misuse, resulting in stricter prescription regulations, as multiple participants mentioned. \textit{"There's a shortage of my ADHD med, and neither of my pharmacies seems to have it, and I'm on my last like, I don't have any left right now,"} expressed P7. 

Getting a doctor or prescription marked only the beginning, particularly for participants grappling with mental health issues. They placed immense importance on not just obtaining medical assistance but on \textbf{finding a trustworthy and empathetic doctor}. P2, for instance, experimented with two therapists before settling on the current one, emphasizing the difficulty in locating a psychiatrist they genuinely trusted and perceived as caring. P4 also remarked: \textit{"It's really hard to find a psychiatrist that you trust, that you think is caring."} The recurrent theme of `caring' resonated strongly throughout the interviews. P4, voicing frustration, explained, 
\begin{quote}
    \textit{"It's not that they couldn't understand what I was saying. They were just like, that's normal, so it doesn't matter, (or:)'I think that has a lot to do with being a woman and being young, and the experiences of people in those demographics within the medical system.'(said by doctor)" (P4)}
\end{quote}
Participants also voiced concerns about doctors overlooking the side effects of medications. One participant lamented, \textit{"Doctors are prescribing. Practitioners are not listening to me when I say that my side effects from this medication are harmful to me, and I do not want to take it anymore."}. Even P5, who had a longstanding trust in their doctor, refrained from discussing concerns about the medication's long-term effects because, during the initial attempt, the doctor "kind of brushed it off a little bit and said I don't have dementia, and it's fine, and that maybe it's just depression." Similarly, P10 shared discomfort with having the same doctor as their mother and found satisfaction in switching to a psychiatric professional who demonstrated genuine care and was willing to adjust the dosage to alleviate symptoms. Given the subjective and self-evaluated nature of mental health diagnoses and treatment, the trust placed in doctors became even more critical. Many participants sought to establish long-term relationships with trusted doctors and hesitated to switch, underscoring the pivotal role of trust in the management of mental health medications. Understanding this dynamic is essential in navigating the complexities of mental health care. 

To summarize, these unique challenges related to various stakeholders hindered students' access to necessary medication, further disrupting their regular medication management routines. Considering these external factors tied to the broader system is crucial when designing technologies to aid in regulating mental health medication.
\section{Discussion} 
\seclabel{discussion}
\subsection{Identify state changes in university students managing mental health medication: Responsibility, Mental status and Self-acceptance}
\review{During our study, we realized that mental health medication management for university students is a very dynamic process. Therefore, it is important to discuss the transitions between different states during the management process.}
\review{In previous work, Caldeira et al. identified four states within the medication management framework: Wellness, New Task, Erratic, and Disruption~\cite{caldeira2022state}. These states were categorized along two dimensions: medication regularity (routine vs. non-routine) and time scale (long-term vs. short-term). The authors stated that changes in routines and medications can disrupt the most stable "Wellness" state. This statement aligns with our findings.  
Based on the interviews, we further classified the specific changes in routines and medications that college students might encounter when managing mental health medications.
For changes in medication, this includes but is not limited to} starting medication, adjusting dosage, and changing the medication. 
\review{For changes in routine, }relocation, and shifts in personal status (examples specific for university students include entering a new school, graduating, moving out from parents, etc.) are special events that college students are facing.
\review{Although the overall categories of factors influencing state changes align with previous work, we do find and want to emphasize some differences. One important observation is that for psychiatric medications, the dynamic transitions are much more frequent, making the 'New Task State' more common than the 'Wellness State'. Additionally, we found that the 'Wellness State' is not always the most desirable state for college students managing their mental health medication. According to our interviews, many participants, despite recognizing the chronic nature of their mental health symptoms, have a strong desire to lower their dosage or stop taking medication altogether. As researchers, it's crucial to acknowledge that the desire to reduce dosage can be an additional cause of states changes, beyond the introduction of new medication as mentioned in previous work.}

\review{In our study, we also want to highlight three additional factors that may contribute to changes in states, beyond routine and medication. 
The first factor is the change in \textbf{Responsibility}. For our participants, who are college students, we found that many are gradually taking on more responsibility for managing their medications independently. This shift in mindset not only causes routine differences, such as the absence of parents' reminders (P1), but also grants them more autonomy in decision-making. This change in responsibility leads to a series of changes in how they manage their medications. 
Therefore, including changes in responsibility within the framework is crucial for thoroughly discussing state transitions.}

\review{\textbf{Mental status} is another factor that can influence medication management states. P5 mentioned that during particularly depressive episodes, even getting out of bed and accessing medication stored in the kitchen became challenging. For our participants, changes in mental status are common due to their diagnoses and the effects of medications. It is important to recognize that for the population we are targeting, fluctuations in mental status can significantly affect how they manage their medications, and should be considered as a common factor to cause state changes.}

\review{Another factor is \textbf{Self-acceptance}. While the stigma associated with mental health disorders may not be as prevalent among university students as in the past~\cite{Gabunia_Verulava_2024}, there remains a hesitancy regarding medication, with many participants not viewing it as the best option. We observed variability in these concerns among our participants. Some mentioned feeling more at peace after several years of medication use. Given that intentional nonadherence remains a significant issue~\cite{lehane2007examination}—especially in our context—it is crucial to include the self-acceptance of medications in our spectrum of considerations to capture possible state changes.}

\subsection{Mutual relationships between mental status/symptoms and medication}
In the earlier discussions, a key point emphasized by our participants is the complex process to \textit{"alter your mental state in an attempt to fix your mental state"}. This intricate relationship makes it challenging to ascertain whether changes in symptoms are attributed to medication or other factors. \review{In many past studies discussing technology developed for medication management, the relationship between medications and mental symptoms was often underexplored, with the focus primarily on the management of medications alone  ~\cite{stawarz2014don}.
For example, in a study by Stawarz et al. reviewing the functionality of 229 medication reminder apps, none of the three design requirements they summarized addressed symptom tracking~\cite{stawarz2014don}.
This is understandable since most chronic diseases are relatively stable, and traditionally, decisions regarding medication dosage and changes are based solely on clinical advice, not self-report. Consequently, the primary responsibility of a patient in managing their medication is simply to "adhere to that medication"~\cite{osterberg2005adherence}.
In studies focusing on psychiatric medications, researchers are beginning to recognize this unique aspects of managing mental health medications related to symptoms, though they still encounter numerous barriers. For example, Papoutsaki et al. explored how individuals in an online health community track their symptoms and psychiatric medication dosages, particularly when attempting to withdraw from a medication. Members of the forums emphasized the use of daily diaries to track symptoms, and some even created charts to monitor the exponential decay of their previous dosage.
However, this approach still heavily relies on self-regulation~\cite{Cadel2021MedicationMF}. In our study, none of our participants used similar tools or social platforms to manage their medication and symptoms.}

\review{To further elaborate on the relationship between medication and mental symptoms, it's important to recognize their interdependence. Tracking changes in symptoms can help determine the effectiveness of the medication and the appropriate dosage. Conversely, fluctuating mental states can also affect a participant's ability to manage their medication. This intrinsic relationship complicates the medication management process. Some of our participants even mentioned managing their medication based on their symptoms rather than adhering to more routinized clinical advice.
Although these actions are less standardized compared to those observed in online communities by Papoutsaki et al., the underlying rationale behind these behaviors is the same~\cite{papoutsaki2021understanding}. The real-world practices we discovered suggest that a standardized medication regimen may not be adequate for participants who are actively engaged in a sensemaking process to understand the impact of their medications on their symptoms.
In this context, technology should be designed to actively assist college students in understanding the relationship between their prescribed medications and their mental symptoms to prevent intentional nonadherence.}

\subsection{Design implications}
\review{Based on our findings in interviews and discussions with previous works, we would like to present several detailed design implications. We hope that those implications can help researchers to design future technology that help university students better mange their medications in six aspects.}
\subsubsection{Tracking the relationship between mental status/symptoms and medication}
\review{First, }the system should have the capability to directly track and display potential relationships between medication and mental health symptoms. 
\review{Other than an application only tracking medication dosage or mental status and mood, the system should proactively establish a relationship between those two factors. Second, it's also essential to} record significant events that may influence mental status during the medication transition process, both those self-identified and automatically tracked. These events include but are not limited to, starting medication, adjusting dosage, changing the medication, relocation, and shifts in personal status (examples specific for university students include entering a new school, graduating, moving out from parents, etc.).
With these functions in place, the system should be able to aid students in evaluating the impact of medication, thereby fostering trust and belief in the treatment. 

\subsubsection{Dynamically adjust reminders based on location and schedule}

We observed that participants typically linked their medication intake with routines and specific locations. In this context, they often encounter dissatisfaction with fixed reminders, particularly when they have flexible routines. This barrier prompts them to establish their own methods, such as creating backup reminders. \review{This finding aligns with previous research; for example, Stawarz et al. mentioned that medication reminder apps need to support the creation of backup reminders and routines~\cite{stawarz2014don}. We aim to further elaborate on these broad design implications drawn from previous works. To design a} future system \review{that} acknowledge and accommodate \review{participants} existing approaches to medication management\review{, we proposed the following actionable design suggestions based on our findings.}\\
\noindent
\textbf{Broad delivery time window based on sleep, eating schedule, last medication intake time, and availability}\indent The reminder's delivery time window should be set widely, especially in situations where medication does not necessitate a fixed intake time (which is the case in 100\% of our interviews). Based on our interviews, we identified \textbf{sleep} and \textbf{eating} activities as two primary factors that can impact the medication-taking process. Therefore, the delivery time of a reminder should be dynamically adjusted according to these activities within this broad window. \textbf{The last medication intake time} is another crucial factor, considering that many medications for mental health conditions have a specific duration of effectiveness, as well as a maximum dosage within a period. If the first dosage is taken too early, the subsequent dosage should not be administered. 
The reminder time should be adjusted accordingly.
After finding a good medication intake time, it's equally important to ensure the reminder is delivered at a time when they can actually take their medication. If the reminder is sent during a meeting or another commitment, they may be unable to take the medication, necessitating additional actions such as setting a backup alarm or even results in forgetting the task eventually. As a result, it is vital to send reminders at times when they are available to take the medications. For university students, academic activities, such as meetings or deadlines, play a significant role in determining their \textbf{availabilities} and thus need to be considered in the decision-making process.

\noindent
\textbf{Deliver at the right place: Storage and activity}\indent During our interviews, using \textbf{location cues} as reminders to take medication was commonly reported. Participants noted that a reminder pop-up on the phone is often ignored in many situations. As a result, the reminders should also be delivered at the right place where they are more likely to be noticed by the participants. These places typically fall into two categories: the location of \textbf{medication storage} and the location of \textbf{certain activities} linked with medication. In some instances, these two places could be the same, for example, keeping medication next to the bed and taking medication before sleep. 
Having reminders delivered close to the medication storage place is also beneficial for easy access to medication right after receiving the reminders, as mentioned in the previous section. For example, an audio reminder could be placed next to a pillbox as participants walked by.
\subsubsection{Recognize the potential for immediate medication intake in response to symptoms}
In our interviews, we also noted that, at times, medication management was loosely based on symptoms. In severe cases, missing a dose and not promptly taking it could lead to a potential disaster (e.g., P7 unable to drive and needing to take the subway home). While the system should provide participants with considerable autonomy to decide in such situations, it's important to acknowledge that participants might take action when faced with these situations, and follow-up management should be adjusted accordingly. Although more discussion is needed, especially considering the limited accuracy of sensing those symptoms and the complicated decision-making process to ensure safety, immediate intervention might be necessary if medication non-adherence occurs in a dangerous situation (e.g., while driving).
\subsubsection{Striking a balance between autonomy and minimal effort}
We recognized a paradox in the medication management process: students desire control over their own routines but are hesitant to invest substantial effort in additional tasks, such as self-reporting. While many of these challenges may arise from limitations in the technology they use\review{~\cite{kelley2017self}}, which could be better designed following the previous design implications, additional design strategies could enhance the overall experience. 

Firstly, automatic tracking and background reminders can significantly alleviate the efforts students invest in the medication management process. However, false detections might lead to increased efforts for correction\review{~\cite{lam2015medication,morisky2008retracted}}. In such instances, these tracking and reminders should be easily modifiable with simple interactions akin to the one-click snooze technique for a phone alarm. This ensures that participants can maintain minimal effort while achieving more accurate and flexible management results.

\subsubsection{Safeguarding privacy in a creative manner}

Many individuals express concerns about third-party applications accessing sensitive information or the potential for unauthorized access. Here are two strategies to alleviate these concerns: 1) Create pseudonyms for medication, especially for reminders displayed on the interface that other people might be able to see. This creative approach was suggested by one of the participants. By integrating a `personal encryption rule' into the process, the application can significantly desensitize the information related to medication management; 2) De-identification, especially regarding academic information. In discussions about collaboration between academic institutions and medical services, despite understanding the regulations, many individuals remain concerned that disclosure of their symptoms, especially medications, might negatively impact their academic well-being. One way to address this concern is to completely de-identify the data (name, email, phone number, academic program). While complete anonymity may be challenging\review{~\cite{walker2016all,Emam2013PrivacyAA}}, implementing some form of redirection in the information transition process can help students build trust in the technology and increase adoption.

\subsubsection{Enhance communication with other stakeholders to build trust}

Many participants conveyed that their doctors did not show sufficient care for their symptoms and side effects. \review{In a paper by Jo et al., the authors discuss how to support clinicians in planning the discontinuation of psychiatric drugs, including the types of information needed~\cite{jo2022designing}. While they considered factors such as trust issues, patient medical history, and insurance issues, they did not address how patient feedback might influence decisions about the regimen. In our study, we found that the relationship between symptoms and medications is crucial for managing mental health medications among university students. This relationship is typically best understood by participants through their daily experiences. Considering the lack of trust and communication expressed by some of our participants, we propose that participants should have the ability to utilize data previously collected on the mutual relationship between symptoms and medications. This would empower them to effectively discuss and influence medical decisions with their doctors.} However, this should be under full control of the participants, taking into consideration privacy issues.

\subsection{Limitations}
This study is subject to certain limitations. Firstly, the sample size is relatively small, restricting our ability to generalize the findings to a wider population. This limitation stems from the reluctance to disclose sensitive information. In our research, we specifically focus on a particular group of university students, requiring them to provide an `edu' email address. Participants may be concerned about linking their academic details to mental health challenges, thereby impeding their willingness to take part. To address these concerns, aside from implementing de-identification measures, we only requested participants to disclose information related to diagnoses and medications if they were comfortable doing so. However, this introduces the potential for selection bias, as those who willingly participated in the interviews might already be predisposed to sharing such information. Despite acknowledging these limitations, our analysis of the interview data revealed numerous common patterns \review{and reached data saturation}, affirming the validity of our findings. Additionally, we provide the survey results encompassing demographic information, medication schedule, medication adherence, and self-efficacy. This information can assist future researchers in better contextualizing our results. \review{Given the diversity in daily routine management, we are cautious about generalizing our findings to other population groups.} Future research could target a broader population using less sensitive measures, such as surveys, to further validate our findings. 

In our study, we refrained from collecting specific medication names for each participant due to privacy considerations. While many participants willingly shared information about symptoms, medication working mechanisms (e.g., immediate effects or delays), or even the exact names of medications, this practice impacted our capacity to systematically analyze varied strategies corresponding to different types of medications. Acting as an exploratory endeavor, we aspire for future studies to gather more comprehensive information, such as electronic health records (EHR), to further delve into these issues.
\section{Conclusion} 
\seclabel{conclusion}
In the paper, we conducted semi-structured interviews with university students currently undergoing medication for mental health issues. The interviews aimed to uncover the challenges they encountered in their university lives, along with the coping strategies and technologies employed to navigate these challenges. We found that deciding to take medications is often a challenging choice for many students. The expectations for dosage reduction, the inherent uncertainty of medication effects, the intricate relationship between medications and symptoms, and the frequent life changes associated with university all contribute to a highly dynamic journey in medication management. As a result, students have developed various strategies to cope with these challenges, often relying on creative ways to use fixed alarms on their phones. However, they encounter numerous difficulties due to factors such as the flexibility of daily routines. Despite university students' strong desire to maintain autonomy in the medication management process, they still struggle to devote additional effort to self-reporting, creating additional challenges in building trust with their doctors. Based on these findings, we proposed six design implications for how advanced technology could better assist university students in managing their mental health medications.




\ifanonymized
 \relax
\else
 \section*{Acknowledgements}
\fi 


\bibliographystyle{plain}	

\bibliography{bibs/local}
\section{Appendix}
%
\begin{longtable}{llll}
\toprule\toprule
Characteristic  &                           & n & \% \\
\endfirsthead
\endhead
\midrule
Gender          &                           &   &    \\
                & Female                    & 8 & 80 \\
                & Male                      & 2 & 20 \\
Age             &                           &   &    \\
                & 18 - 24                   & 4 & 40 \\
                & 25 - 34                   & 6 & 60 \\
Race/Ethnicity  &                           &   &    \\
                & White                     & 5 & 50 \\
                & Hispanic                  & 1 & 10 \\
                & Black or African American & 1 & 10 \\
                & Asian                     & 1 & 10 \\
                & Other                     & 2 & 20 \\
Education level &                           &   &    \\
                & Some college              & 3 & 30 \\
                & Professional degree       & 4 & 40 \\
                & Doctorate                 & 3 & 30 \\
Marital status  &                           &   &    \\
                & Never married             & 8 & 80 \\
                & Married                   & 1 & 10 \\
                & Other                     & 1 & 10 \\
Income          &                           &   &    \\
                & Less than \$10,000        & 3 & 30 \\
                & $50,000 - $59,999         & 3 & 30 \\
                & $50,000 - $59,999         & 3 & 30 \\
                & Blank                     & 1 & 10 \\

\bottomrule\bottomrule

\label{tab:my-table}\\
\caption{Demographic information of the participants in number and percentage (n=10)}
\end{longtable}

\begin{longtable}{llllll}
\toprule\toprule
Characteristic          &                                 & n  & \% & Mean & SD   \\
\endfirsthead
\endhead
\midrule
Diagnosis length        &                                 &    &    &      &      \\
                        & 1-3 years                       & 2  &    &      &      \\
                        & 3-5 years                       & 4  &    &      &      \\
                        & more than 5 years               & 4  &    &      &      \\
Medication length  &&&&&\\&  1-3 years                       & 6  &    &      &      \\
                        & 3-5 years                       & 0  &    &      &      \\
                        & more than 5 years               & 1  &    &      &      \\
                        & Not answered                    & 3  &    &      &      \\
\begin{tabular}[t]{@{}l@{}l@{}}
Total medication \\counts
\end{tabular} &                                 &    &    & 1.6  & 0.70 \\
Frequency               &                                 &    &    &      &      \\
                        & Daily (once a day)              & 10 &    &      &      \\
                        & Weekly (once a week)            & 1  &    &      &      \\
                        & Multiple times a day            & 1  &    &      &      \\
Pill containers         &                                 &    &    &      &      \\
                        & Original pill bottle            & 8  &    &      &      \\
                        & Original blister                & 1  &    &      &      \\
                                             & 7-day drug store like pillbox or similar one                                         & 2 &  &      &      \\
Obtain medication       &                                 &    &    &      &      \\
                        & On-campus healthcare/clinic     & 1  &    &      &      \\
                        & Off-campus healthcare provider  & 5  &    &      &      \\
                        & Local pharmacy/chemist          & 2  &    &      &      \\
                        & Online pharmacy services        & 1  &    &      &      \\
                        & Other                           & 1  &    &      &      \\
\begin{tabular}[t]{@{}l@{}l@{}}
Medication management \\strategics
\end{tabular}            &&&&&\\& Establish a stable routine/schedule                             & 8 &  &      &      \\
                        & Digital reminders/alarms        & 6  &    &      &      \\
                        & Physical reminders/ visual cues & 5  &    &      &      \\
                        & Self-report on digital devices  & 2  &    &      &      \\
                        & Families/friends remind me      & 2  &    &      &      \\
                        & No, I depend on my memory       & 1  &    &      &      \\
Technology used         &                                 &    &    &      &      \\
                                             & \begin{tabular}[t]{@{}l@{}l@{}}
Digital calendar/ reminder APPs \\(phone calendar, google calendar,\\phone alarm, etc.)
\end{tabular} & 6 &  &      &      \\
                                             & Third-party medication reminding APPs                                                & 1 &  &      &      \\
                                             & Digital self-report APPs (note, diary APP, etc.)                                     & 1 &  &      &      \\
                        & No, I depend on my memory       & 4  &    &      &      \\
\begin{tabular}[t]{@{}l@{}l@{}}
SEAMS self-efficacy\\(non 13- efficacious 39)
\end{tabular}&                                                                                      &   &  & 25.3 & 4.74 \\
\begin{tabular}[t]{@{}l@{}l@{}}
MMAS-8 adherence\\(adhere 0- non 7)
\end{tabular}         &                                                                                      &   &  & 4.2  & 1.87 \\
\bottomrule\bottomrule                              
\label{tab:survey1}\\
\caption{Baseline information related to mental health and medication}
\end{longtable}

\begin{longtable}{lll}
\toprule\toprule
Question (never 0 - always 4) & mean & sd \\
\endfirsthead
\endhead
\midrule
How often do you have difficulty remembering to take all your medication?                 & 1.8  & 1.03 \\
\begin{tabular}[t]{@{}l@{}l@{}}
Do you think your mental health status affects your adherence to your\\medication? \end{tabular}        & 1.86 & 1.57 \\
\begin{tabular}[t]{@{}l@{}l@{}}After missing one medication, will you be more conscious of taking your\\next dose?\end{tabular}        & 3    & 1    \\
\begin{tabular}[t]{@{}l@{}l@{}}After missing one medication, how often do you continue to miss several \\subsequent doses?\end{tabular} & 0.86 & 0.69 \\
\bottomrule\bottomrule
\label{tab:survey2}\\
\caption{Answers of the self designed survey}
\end{longtable}

\end{document}